\documentclass[useAMS,usenatbib]{mn2e}
\newcommand \mnras{{MNRAS}}
\newcommand \apj{{ApJ}}
\newcommand \aj{{AJ}}
\newcommand \apjs{{ApJS}}
\newcommand \pasp{{PASP}}

\newcommand \aap{{A\&A}}

\usepackage{epsfig}
\usepackage{natbib}
\usepackage{subfigure}
\usepackage{amssymb}
\def \Msolar {$M_{\odot}$}
\def \Mstar {$M_{\star}$}
\def \Critmass {$\sim 10^{10}$ \Msolar}
\def \Critmassel {$\sim 10^{11}$ \Msolar}

\def \magarc {mag arcsec$^{-2}$}

\begin{document}

\title[Are bulges simply spheroids wrapped in discs?]{The signature of dissipation in the mass-size relation: are bulges simply spheroids wrapped in a disc?}
\author[Berg et al.]{Trystyn A. M. Berg$^1$, Luc Simard$^{2}$, J. Trevor Mendel$^{3}$, Sara L. Ellison$^{1}$\\
$^1$Department of Physics and Astronomy, University of Victoria, Victoria, BC, V8W 2Y2, Canada \\
$^2$National Research Council of Canada, 5071 West Saanich Road, Victoria, BC, V9E 2E7, Canada \\
$^3$Max Planck Institut f\"ur Extraterrestrische Physik, Giessenbachstra\ss{}e, D-85748 Garching, Germany }

\maketitle

\begin{abstract}
The relation between the stellar mass and size of a galaxy's structural subcomponents, such as discs and spheroids, is a powerful way to understand the processes involved in their formation. Using very large catalogues of photometric bulge+disc structural decompositions and stellar masses from the Sloan Digital Sky Survey Data Release Seven, we carefully define two large subsamples of spheroids in a quantitative manner such that both samples share similar characteristics with one important exception: the `bulges' are embedded in a disc and the `pure spheroids' are galaxies with a single structural component. Our bulge and pure spheroid subsample sizes are 76,012 and 171,243 respectively. Above a stellar mass of \Critmass{}, the mass-size relations of both subsamples are parallel to one another and are close to lines of constant surface mass density. However, the relations are offset by a factor of 1.4, which may be explained by the dominance of dissipation in their formation processes. Whereas the size-mass relation of bulges in discs is consistent with gas-rich mergers, pure spheroids appear to have been formed via a combination of `dry' and `wet' mergers.
\end{abstract}


\section{Introduction}

The structure of galaxies emerges from a complex process of hierarchical mass assembly in which mergers and other processes such as dissipation play key roles. Discs and spheroids contain the bulk of the light (and thus stellar mass) of galaxies, and their scaling relations encode valuable information about their formation and evolution \citep[e.g.][]{fall80, djorgovski87, burstein97, mo98}. Discs and spheroids have been most often distinguished on the basis of photometric data from which light profiles are measured and morphological decompositions are performed \citep[e.g.][]{freeman70,boroson81,kent85,dejong96,peng02,simard02,simard11,lackner12,meert13}. Under this observational classification paradigm, spheroids are parametrically described by S\'ersic/de Vaucouleurs light profiles, whereas discs' profiles follow an exponential decline. However, this relatively simple picture is changing with the greater availability of detailed internal kinematics data, e. g., ATLAS$^{\rm 3D}$\citep{krajnovic12}. Through comparisons between photometric and kinematical data, fast rotators have been associated with disc galaxies and slow rotators with dispersion-supported spheroids. The fast/slow rotator classification is also closer to the definition of structure in numerical simulations \citep[e.g.][]{abadi03,governato07,hopkins09a,agertz11,brooks11,scannapieco11}. High angular momentum particles in these simulations are associated with a disc, and spheroids comprise the low angular momentum particles. 

An extensive body of theoretical and observational studies \citep[e.g.,][]{bender92,ciotti01,shen03,robertson06,ciotti07,scarlata07,hopkins09a,nipoti09,auger10,nipoti12,shankar13,taranu13} has shown that merger histories and hydrodynamical processes such as dissipation shape the properties of spheroids. Much of this work has focused on reproducing the mass-size relation. This fundamental relation describes how the sizes of galaxies scale with their stellar mass. The relation is a projection of the Fundamental Plane \citep{djorgovski87}, an inherent relation for all dispersion supported galaxies and is typically very tight \citep[e.g.][]{lauer07,kormendy09}.  However, there are interesting trends that remain to be studied in greater detail. For example, \citet{shen03} found that the merger of two similar late-type galaxies produced an apparent flattening of the mass-size relation of the remnant spheroid and that disc instabilities were responsible for growing the bulge. \citet{shankar13} found a similar flattening, but they attributed it to the lack of gas in the merging galaxies. There is also evidence that bulges are not present-day elliptical galaxies embedded in discs \citep{graham08,gadotti09,gadotti09-2,laurikainen10} as they exhibit different mass-size relations. 

The gas content of a merger progenitor can determine the size of the virialized remnant as it controls the amount of energy that can be lost as a result of dissipation. Gas-rich (`wet') mergers tend to dissipate orbital energy, which shrinks the effective size of the system as the gas cools. The light profile of such a galaxy will have a higher stellar surface density due to the increasing central concentration of stars \citep[][and references therein]{hopkins09b}. However gas-poor (`dry') mergers maintain this energy in stellar orbits, keeping the galaxy puffed up. Violent relaxation can mimic dry mergers by ejecting stars into the outer regions of galaxies, making the light profile extend to larger radii. Discs can regrow and be replenished around these merger remnants both by diffuse infall from the surroundings and by gas accreted from satellite galaxies \citep[e.g.][]{guo11}. Although one might expect discs to be destroyed in the merging process itself, both \citet{robertson06} and \citet{hopkins09b} (amongst others) demonstrated that discs can quickly reform during the merger with sufficiently gas-rich progenitor galaxies. The bulge is then built very slowly with little angular momentum, keeping it small. Finally, the bulge component of disc galaxies can also form via secular instabilities in the disc, transferring some disc material into the bulge component \citep[][and references therein]{courteau96, shen03}.

\citet{shankar13} have recently demonstrated with semi-analytic models of galaxy evolution \citep{guo11} that the mass-size relation flattens for spheroidal galaxies formed by \emph{dissipationless} processes below a critical mass of \Critmassel{}. This result differs from their sample of observations of early-type galaxies from the Sloan Digital Sky Survey (SDSS). The pure power law description of their data was suggested to be the result of \emph{dissipation} of energy due to wet mergers. The flattening effect seen below a critical mass is believed to be an event of dry mergers as the excess energy puffs up the spheroidal component of the galaxy.  \citet{hopkins09b} also simulated several mergers of different masses and different gas contents. They found that when dry mergers are turned off (i.e. only wet mergers); the sizes are much smaller than expected, although the mass-size relation still remains more or less a power law. When gas was treated like stars (i.e. no dissipation), a power law similar to the `no dry merger' case resulted; however, these sizes are larger than those from wet mergers. These two studies show that the size of the bulges in gas-rich galaxies seems to be smaller than their dry-merging counterparts. However, both pieces of work disagree on how strong the effects of dissipation are as their mass-size relations are inconsistent with each other, especially with the flattening seen by \citet{shankar13} at low masses.

In order to better understand the effects of dissipation on the formation of the spheroidal components of galaxies, we study the mass-size relations of two very large spheroidal subsamples, namely bulges and pure spheroids, and compare them with expectations from theoretical models. The careful, quantitative selection of our subsamples, derivation of spheroid masses and the characterization of the mass-size relation are described in Section~\ref{data}.  Our results are discussed in Section~\ref{results}. The cosmology adopted throughout this paper is ($H_0, \Omega_{m}, \Omega_{\Lambda}$) = (70 km s$^{-1}$ Mpc$^{-1}$, 0.3, 0.7).


\section{Data}\label{data}

\subsection{Sample Selection}
The taxonomy of describing galaxy subcomponents in the literature is quite diverse. In addition to galaxy discs, the literature is filled with pseudo-bulges, classical bulges, lenticulars, and spheroidals just to name a few \citep{kormendy12}. We adopt ``spheroid" here to describe the structural component of a galaxy that is not a disc (and not a bar). Spheroids can then be divided into two categories: bulges which are spheroids embedded in discs and pure spheroids which correspond to single component galaxies. We seek to quantitatively define bulge and pure spheroid subsamples by applying a set of selection criteria to the large catalogues of $n=4$ bulge+disc decompositions and stellar masses presented by \citet{simard11} (S11 hereafter) and \citet{mendel14} (M14 hereafter) for spectroscopically-observed galaxies in the Legacy area of the SDSS Data Release Seven \citep{abazajian09}. Even though the parameters of the S11 decompositions do match those measured from deeper, higher resolution observations \citep[e.g.][]{Kanwar08}; the spatial resolution of SDSS images is unfortunately insufficient over the magnitude range of our sample to accurately measure the S\'ersic index of the bulge itself. We therefore cannot differentiate between different kinds of bulge profiles. Furthermore, our photometric decompositions cannot differentiate between virially-supported or disc-like \citep[e.g.][]{cappellari11} structures without any kinematic data.

We start by selecting galaxies with effective $r$-band surface brightness $\mu_{50,r} \leq 23.0$ \magarc{}, SDSS data base flag {\tt Specclass} = 2 and spectroscopic redshifts in the range $0.005 \leq z \leq 0.2$. The cut on surface brightness was necessary as fainter galaxies up to $24$ \magarc{} were only targeted when the global and local sky values were within 0.5 mag of each other. The {\tt Specclass} flag selects objects that have been identified as galaxies on the basis of their spectral energy distribution. The redshift selection ensures that peculiar velocities do not dominate the distance determination of galaxies in our sample. We applied a quality control criterion to our SDSS catalogues. We used $\Delta_{B+D}$ to keep only galaxies for which the sum of their individual bulge and disc masses is within $3\sigma$ of their total stellar mass as a consistency check (as discussed in M14).

M14 discuss the classification of galaxies based on their 1D profiles into one of four types: (1) those dominated by the bulge component, (2) those dominated by a disc component, (3) those which host a classical bulge+disc (`B+D') configuration, and others that do not fit these categories.  For the purposes of this work we rely on profile types 1 and 3. For a bulge within a disc, we require that the galaxy be classified as a profile type 3. However, it is unclear in the profile fitting whether a pure spheroid system is best fitted by either a single bulge component (type 1) or B+D  configuration (type 3). To ensure we are correctly selecting the single-component spheroids and not including the B+D systems, we use the $F$-test probability ($P_{pS}$) calculated in S11 which indicates whether a given galaxy is better fitted with the pure S\'ersic (`pS') model rather than the B+D model. This image modelling was originally performed using the SDSS $gr$ images but has since been extended to the $ugriz$ filter set to compute galaxy, bulge and disc stellar masses as described in M14.  Therefore, the distinction between pure spheroids and bulges is made on the basis of $P_{pS}$, the profile type, and the $r$-band bulge fraction ($(B/T)_r$). As shown in figure 12 of S11, the combined criteria $0.0 \leq P_{pS}<0.32$ and $(B/T)_r > 0$ select galaxies for which two-component fitting is required, and the combined criteria $0.32\leq P_{pS}\leq 1.0$ and $(B/T)_r > 0.5$ select single-component galaxies. The profile type further refines the sample and ensures that  
the $P_{pS}$ cut is indeed a physical representation of the light profile.

The remaining three selection criteria are disc inclination $i$, ellipticity $e$ ($e \equiv 1-\frac{b}{a}$) and effective radius $R_{e,kpc}$. Internal extinction can have a significant impact on measured structural parameters \citep{pastrav13}. Given that our bulge subsample was still quite large at this point, we opted to keep only bulges in face-on ($i \leq 60^{o}$) galaxies to improve the quality of this subsample. One difficulty with bulge+disc decompositions of galaxy images is that strong bars can be mistaken for bulges. The signature of this problem is a very high bulge elllipticity. Selecting galaxies with $e < 0.6$ is sufficient to remove bars from the bulge subsample. Finally, we required that the bulge effective radius be smaller than the disc. Our selection criteria\footnote{These criteria do not introduce any differences in the magnitude distribution between the two subsamples.} are summarized in Table~\ref{tab:cuts}.

\begin{table}
\caption{Selection criteria and subsample sizes.}
\begin{center}
\begin{tabular}{lcccc}
\hline
Quantity & pS & B+D & N(pS) & N(B+D) \\
\hline
S11 sample &\multicolumn{2}{c}{n/a} & 1,123,718 & 1,123,718\\
In M14 & \multicolumn{2}{c}{n/a} & 656,797 & 656,797\\
$\mu_{50,r}$ &  \multicolumn{2}{c}{$\leq 23.0$ \magarc{} }  & 654,306 & 654,306\\
$specclass$ & \multicolumn{2}{c}{2} & 652,748 & 654,306\\
$z$ & \multicolumn{2}{c}{0.005 -- 0.2} & 615,719 & 617,264 \\
$prcflag$ &\multicolumn{2}{c}{0} & 615,719 & 617,264 \\
$\Delta_{B+D}$ & \multicolumn{2}{c}{$<3\sigma$} & 613,661  & 615,203\\
$P_{pS}$ & 0.32 -- 1.0 & 0.0 -- 0.32 & 429,418 & 185,034\\
Profile Type & 1 or 3 & 3 & 341,993 & 170,544 \\
$(B/T)_{r}$ & $>0.5$ & $>0.0$ & 187,227  & 170,544\\
$i$ &n/a & $\leq60^{\circ}$ & 187,227 & 104,218\\
$R_{e, kpc}$ & $>0.0$ & $<1.67 R_{d,kpc}$ & 187,227 & 104,065\\
$e$ & $<0.6$ & $<0.6$ & 171,243 & 76,012\\
\hline
\end{tabular}
\label{tab:cuts}
\end{center}
\end{table}

\subsection{\label{sec:anl}Analysis}

We build our mass-size relations using the stellar mass (\Mstar{}), the maximum volume sampled ($V_{max}$), and the bulge or spheroid radius $R_{e}$ in kiloparsecs. The radii given in S11 are measured along the semimajor axes of the galaxies. We circularize these radii using the well-known relation

\begin{equation} 
R_{e, circ}= R_{e,sma}\sqrt{1-e},
\label{eq:rcirc}
\end{equation} 

\noindent $R_{e, circ}$ is hereafter simply denoted $R_e$. The stellar masses are taken from M14. They are estimated from the comparison of the galaxies' \emph{ugriz} spectral energy distributions with a large grid of photometric templates which encompasses a range of ages, metallicities, star-formation histories and dust contents. This grid is generated from the Flexible Stellar Population Synthesis code \citep{conroy09} in conjunction with a Chabrier initial mass function \citep{chabrier03}. This methodology was applied independently to both bulges and discs, deriving the masses from the flux in the regions identified by the S11 decompositions. The errors are typically $\sim$ 0.15$-$0.2 dex, and include the statistical uncertainties from the total fluxes and bulge-to-total ratios measured by S11.

\begin{figure}
\begin{center}
\includegraphics[width=0.5\textwidth]{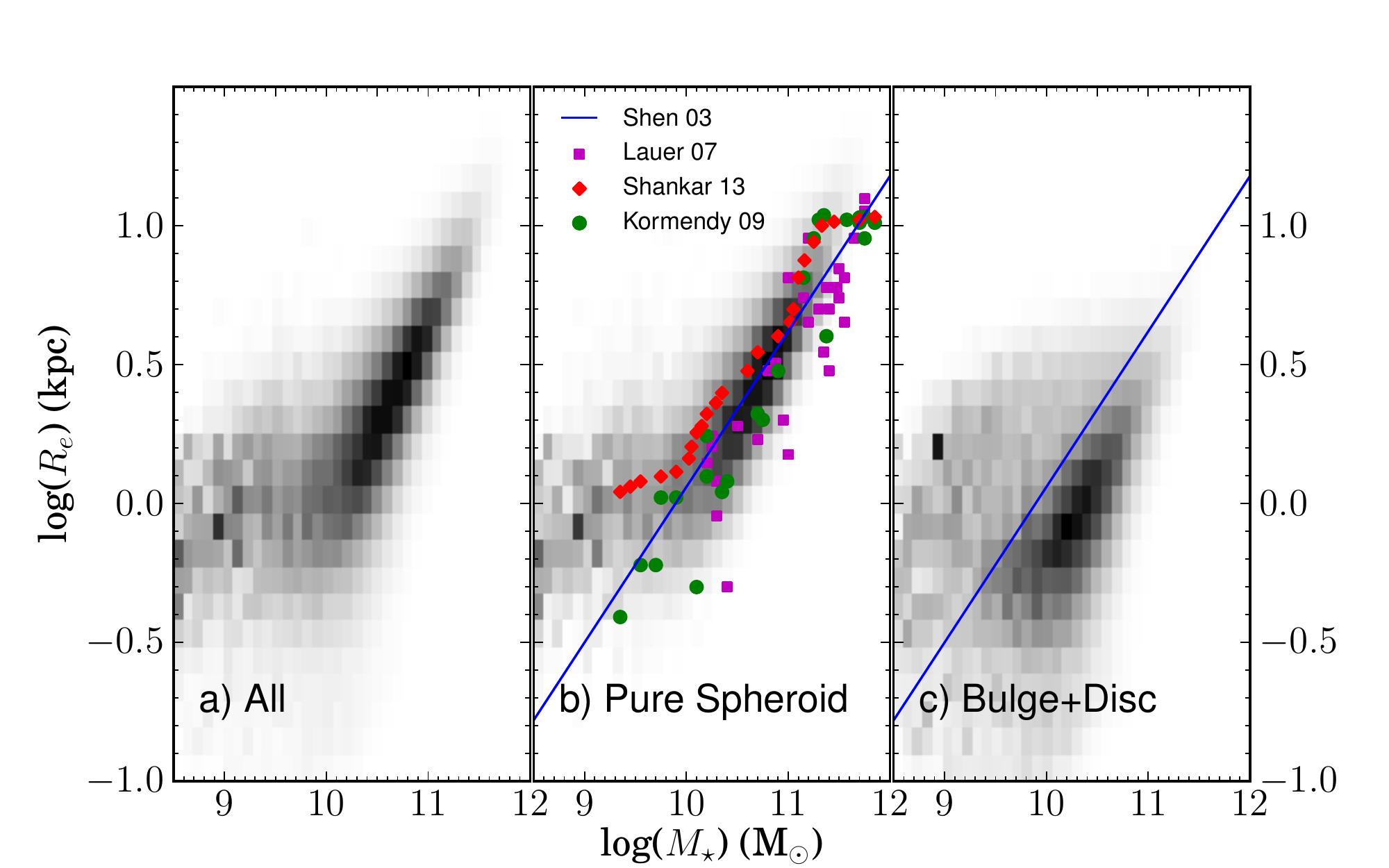}
\caption{\label{fig:grscl} The mass-size relation for (a) the entire sample; (b) the B+D subsample; and (c) the pure spheroid subsample using the $V_{max}$-corrected space density of our subsamples. Data from a number of previous works \citep{shen03,lauer07,kormendy09,shankar13} are overplotted in the middle panel. Stellar masses for the \citet{lauer07} and \citet{kormendy09} data are adopted from \citet{hopkins09b}. The lines from \citet{shen03} represent their best fitting curve to their early-type data set, while the \citet{shankar13} points represent the median of the distribution of their early-type galaxies.}
\end{center}
\end{figure}

Figure \ref{fig:grscl} shows the distributions in the mass-size plane of the entire sample, the B+D subsample, and the pS subsample. Both subsamples show a well-defined trend in the mass-size relation, with little scatter (in particular the pS subset). However, there is a large amount of scatter at the low mass end in the B+D subsample. It is apparent that both subsamples do \emph{not} have similar trends, as the majority of B+D systems tend to have smaller sizes for a given mass than their pS counterparts. To compare to previous work, the observed mass-size data from \citet{shen03}, \citet{lauer07}, \citet{kormendy09}, and \citet{shankar13}\footnote{We adopt the stellar masses derived in \citet{hopkins09b} for the \citet{lauer07} and \citet{kormendy09} data.} have been included in the middle panel of Figure \ref{fig:grscl}. Although the distributions do not line up exactly, it is apparent that the spheroid subsample is very similar to the previous measurements. We see the same flattening in the bulge mass-size relation at low masses as reported in \citet{shankar13}.

\begin{figure*}
\includegraphics[width=\textwidth,angle=0]{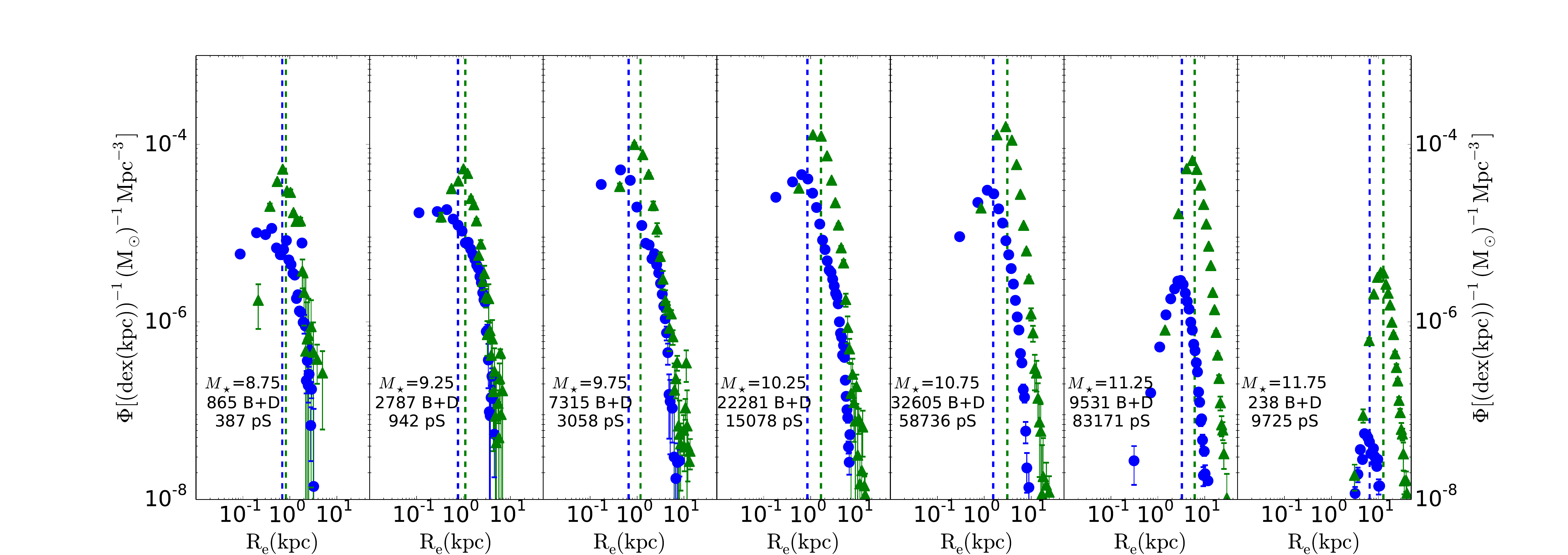}
\caption{\label{fig:hist} Size functions in bins of spheroid stellar mass. The green triangles and blue circles are the spheroid and bulge subsamples, respectively. The centre of the mass bin is displayed at the bottom of each panel, along with the total number of galaxies in each subsample. The error on each point is generated via a bootstrap method. The vertical dashed lines mark the $V_{max}$-weighted median values of the distributions. }
\end{figure*}

In order to characterize the mass-size relation, both subsets were split into seven equally spaced stellar mass bins, and the size functions of each bin were computed and are shown in Figure \ref{fig:hist}.  We calculate the $V_{max}$-weighted median size\footnote{Each galaxy was volume-weighted by $V_{max}^{-1}$  to account for the fact that more luminous galaxies are over-represented in magnitude-limited samples compared to fainter ones.} (where $V_{max}$ is the maximum volume over which a galaxy can be observed from the M14 catalogues) for each mass bin.  Although the scatter for bulges in the lower mass bins is not quite Gaussian, with an extra tail at smaller radii, a Gauss-Hermite expansion of the size functions \citep[e.g., ][]{abraham95} yields mean sizes that agree with the medians to within 0.05 dex.

\section{Results and Discussion}\label{results}

\begin{figure}
\includegraphics[width=0.5\textwidth]{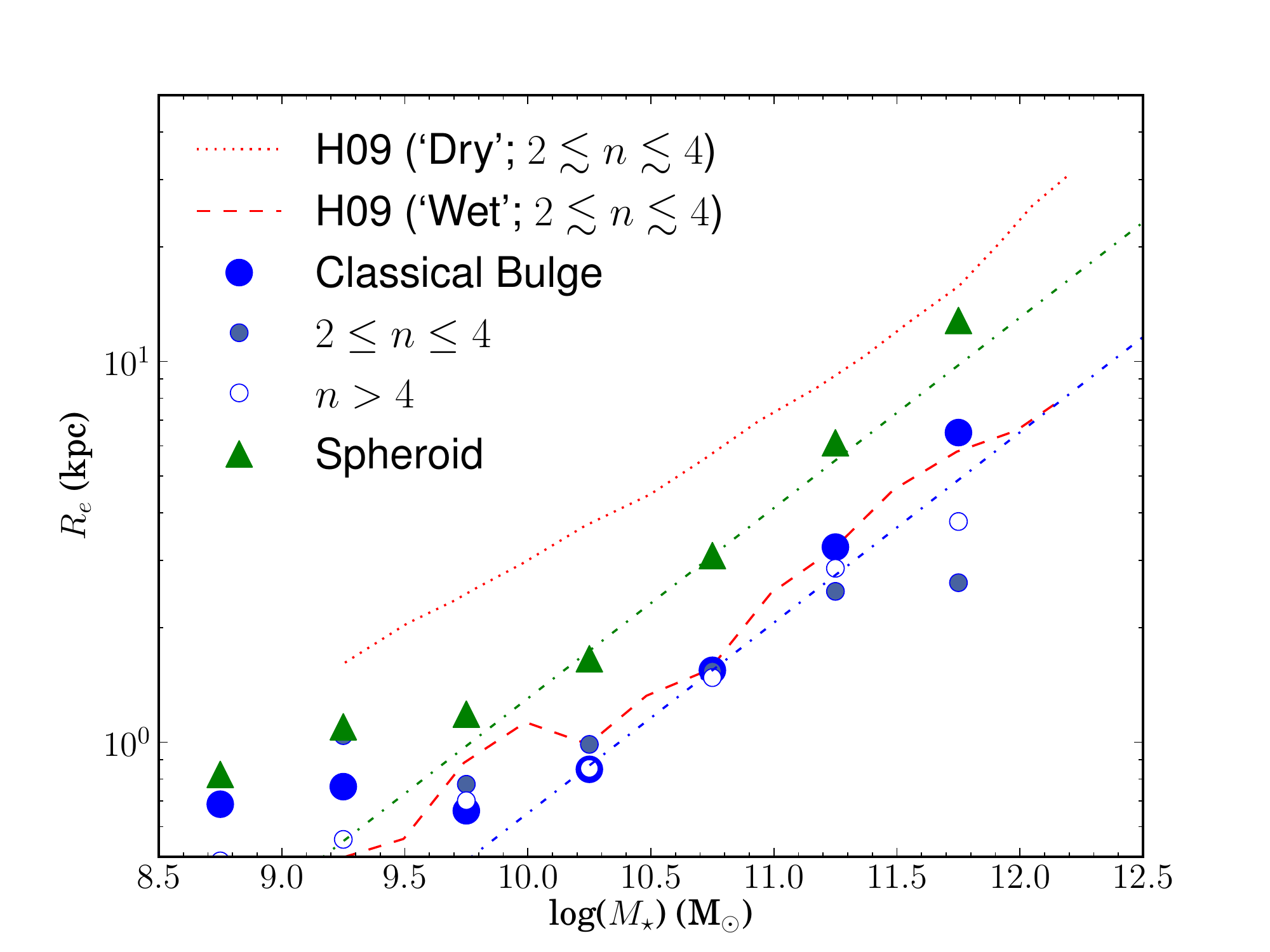}
\caption{\label{fig:mass-size}Stellar mass-size relations for bulges (blue circles) and spheroids (green triangles) using the median sizes shown in Figure \ref{fig:hist}.  The smaller circles represent the mass-size relation using different cuts in the S\'ersic index of the bulge. Due to the large number of galaxies within each bin, the standard errors are much smaller than the data points and thus are not plotted. The `no dissipation' (or `dry'; dotted line) and `no dry' (or `wet'; dashed line) merger models from \citet{hopkins09b} are shown by red lines, for which galaxies typically have $2\lesssim n \lesssim 4$. For masses above \Critmass{}, the slope of the pure spheroid mass-size relation seems to follow models that do not include dissipation, whereas the B+D galaxies (regardless of $n$) follow models without dry mergers or include dissipation. The green and blue dash-dotted lines represent  lines of constant surface density of stars (${\rm log}(R_{e}) \propto 0.5{\rm log}(M_{\star})$) with $\Sigma = 5.9\times10^{9}$ and $2.4\times10^{10} M_{\odot}$ kpc$^{-2}$ respectively.}
\end{figure}

Using the medians from the size functions (Figure \ref{fig:hist}), the bulge and pure spheroid  mass-size relations are shown in Figure \ref{fig:mass-size}.  For comparison, two different models from \citet{hopkins09b} have also been included\footnote{The `no dissipation' and `no dry' models from \citet{hopkins09b} are relabelled as `dry' and `wet' (respectively) to clarify which merger types are dominating in the models.}. These models investigate the predicted mass-size relation at $z = 0$ by: (i) excluding mergers with dissipation (treating gas particles like stars), and (ii) excluding dry mergers. These two extreme cases produce size mass relations that are parallel to one another, but where the sizes of the bulges in dry mergers are larger than in the gas-rich case.  The existence of two distinct mass-size relations in the data, that are also parallel, yet offset, is strong evidence that bulges and pure spheroids form from different types of mergers. {\it Bulges are not simply pure spheroids wrapped in discs}. This is in agreement with previous studies \citep{gadotti09,gadotti09-2,laurikainen10}, although our work has a more rigorously defined sample of bulges and spheroids and exceeds the sample sizes of previous works by at least an order of magnitude.

The \cite{hopkins09b} models show a large spread in the S\'ersic index (predominantly between $2\lesssim n\lesssim 4$) for the progenitor bulge-like structures. To test the effects of assuming a classical B+D decomposition over a free S\'ersic index+disc profile, Figure \ref{fig:mass-size} shows the mass-size relation for bulges with S\'ersic within two bins ($2\leq n\leq4$ and $n>4$; blue grey and unfilled circles, respectively) from the S11 catalogue. The masses were derived using the g-r colours in \cite{dutton09} as M14 did not derive masses for these free $n$ fits. Overall, there is no great difference between assuming $n=4$ or adopting a best fitting $n$ for the bulge\footnote{Despite the apparent disagreement between the mass-size relations for the classical bulge and free $n$ fits for the highest mass bin in Figure \ref{fig:mass-size}, the points agree within the 1$\sigma$ range.} with this data set, therefore we continue to use $n=4$ to represent the populations of bulges.

The first remarkable feature of these relations is their behaviour at masses higher than \Critmass{}. The two relations are clearly shifted with respect to each other, with pure spheroids being 1.4$\times$ larger than bulges at a fixed stellar mass. This offset is so constant over the mass range $10^{10} M_\odot < M_\star < 10^{12} M_\odot$ that both relations would lie directly on top of each other if this shift were applied to the bulges.  They almost follow lines of constant stellar mass surface density (${\rm log}(R_{e}) \propto {\rm 0.5log}(M_{\star})$) in contrast to the virial theorem, which at fixed velocity dispersion, would have ${\rm log}(R_{e}) \propto {\rm log}(M_{\star})$ . The `no-dry' (`wet') merger model of \citet{hopkins09b} reproduces the mass-size relation of bulges very well, but neither model is a good match  overall to the pure spheroid mass-size relation. The observed relation of pure spheroids approaches the `no dissipation' (`dry') model at the highest stellar masses but falls significantly below it for log $M_{\star} < 11.5$ . Clearly, the process that led to the formation of the pure spheroids was not entirely dissipationless, and was therefore likely a mix of wet and dry mergers. This is consistent with \citet{nipoti12} who found that present-day, massive early-type galaxies could not have formed entirely through dry merging. Indeed,  \citet{Kaviraj12} and \citet{Shabala12} have recently identified a specific sample of elliptical galaxies with enhanced star formation rates and nuclear activity that they propose to be the result of a merger.

The second notable feature of these relations is the flattening of size with decreasing mass below \Critmass{}. This flattening was also observed by \citet{shankar13}, and, although they attributed this flattening to a lack of dissipation, none of the Hopkins et al. model curves reproduces this behaviour well. Bulge formation through secular evolution is expected to play a larger role at lower bulge masses and may be responsible for this flattening of the bulge mass-size relation, but it obviously cannot affect the pure spheroid relation in any way. Either dry mergers must dominate the formation of low mass pS galaxies \citep[][]{shankar13}, or another process must be taking place. 

Our bulge and pure spheroid mass-size relations and their observed features point towards a picture in which spheroids do emerge from a hierarchical mass assembly process driven by mergers. However the amount of dissipation, i.e., the amount of gas present in these mergers, leaves a distinctive signature on their present-day sizes and masses.


\section{Conclusion}
Using large catalogues of bulge+disc structural parameters and stellar masses for galaxies in the SDSS DR7 Legacy Area, we quantitatively defined two subsamples of galaxy spheroids to distinctly select pure spheroids and bulges with identical properties other than bulges being embedded in disks. These two subsamples are used to study the relation between their size and stellar mass. We found that these subsamples had distinct mass-size relations, which indicates that bulges are not simply present day, pure spheroids wrapped in discs. At stellar masses above \Critmass{}, the relations are parallel but shifted by a factor of 1.4 in size at a fixed stellar mass. Bulges follow a mass-size relation consistent with a `wet' merger model, and the relation for pure spheroids indicates a combination of `wet' and `dry' mergers. At stellar masses below \Critmass{}, both relations exhibit a flattening in size with decreasing mass that is not reproduced at all by theoretical models. A comparison between our observed relations and theoretical models suggest that dissipation leaves a distinctive signature on the mass-size relations of bulges and pure spheroids.

\section*{Acknowledgements}
We thank Philip Hopkins for graciously providing us with his merger models.

\end{document}